\documentclass[conference]{IEEEtran}
\usepackage{cite}
\usepackage{amsmath,amssymb,amsfonts}
\usepackage{algorithmic}
\usepackage{graphicx}
\usepackage{textcomp}
\usepackage{xcolor}

\usepackage{siunitx}
\usepackage{multirow}

\def\BibTeX{{\rm B\kern-.05em{\sc i\kern-.025em b}\kern-.08em
    T\kern-.1667em\lower.7ex\hbox{E}\kern-.125emX}}

\begin{document}

\title{Movement Detection of Tongue and Related Body Parts Using IR-UWB Radar}

\author{\IEEEauthorblockN{Sunghwa Lee}
\IEEEauthorblockA{\textit{School of Integrated Technology} \\
\textit{Yonsei University}\\
Incheon, Korea \\
sunghwa.lee@yonsei.ac.kr} 
\and
\IEEEauthorblockN{Younghoon Shin${}^{*}$}
\IEEEauthorblockA{\textit{Robotics Lab} \\
\textit{Hyundai Motor Company}\\
Uiwang, Korea \\
yh.s@hyundai.com} 
{\small${}^{*}$ Corresponding author}
}

\maketitle

\begin{abstract}
Because an impulse radio ultra-wideband (IR-UWB) radar can detect targets with high accuracy, work through occluding materials, and operate without contact, it is an attractive hardware solution for building silent speech interfaces, which are non-audio-based speech communication devices. As tongue movement is strongly engaged in pronunciation, detecting its movement is crucial for developing silent speech interfaces.
In this study, we attempted to classify the motionless and moving states of an invisible tongue and its related body parts using an IR-UWB radar whose antennas were pointed toward the participant's chin. 
Using the proposed feature extraction algorithm and a Gaussian mixture model--hidden Markov model, we classified two states of the invisible tongue of four individual participants with a minimum accuracy of 90\%. 
\end{abstract}


\section{Introduction}

Radio frequency electromagnetic waves can be utilized for a variety of applications \cite{Lee2020939, Shin2017617, Lee20191187, Lee22:Nonlinear, Lee2020:Preliminary, Kim22:First, Kang21:Indoor, Lee22:Evaluation, Lee22:Urban, Lee22:SFOL, Rhee21:Enhanced, Yoon14:Medium, Park2021919, Park21:Indoor, Jeong2020958, Lee22:Optimal, Kim21:GPS, Jeong21:Development, Jia21:Ground, Kim2020796, Park2020824, Lee20202347, Kang20191182, Son20191828}.
A radar is a system or sensor that utilizes radio frequency electromagnetic waves to determine various characteristics of objects, such as position, shape, and motion information. Conventionally, radar technology has mainly been applied to long-range applications (i.e., detecting objects from over several hundreds of meters), such as weather forecasting \cite{Doviak1979, Serafin2000}, air traffic control \cite{Shrader1973}, and missile defense \cite{Camp2000}. However, as miniaturized and low-cost commercial off-the-shelf radar appears, radar technology has recently expanded to short-range applications (i.e., detecting objects at a distance less than several meters or even less than a meter). 
Human--computer interaction (HCI) \cite{Moon22:Fast, Moon21, Moon202013, Moon22:Speeding, Moon2019157, Moon2019258} is a prevalent short-range radar-based research field. Indeed, owing to some properties, including high precision, robustness against light changes, and capability to detect targets through materials, radars are a promising sensor solution for HCI \cite{Jaime2016}.

 \begin{figure}
  \centering
  \includegraphics[width=0.9\linewidth]{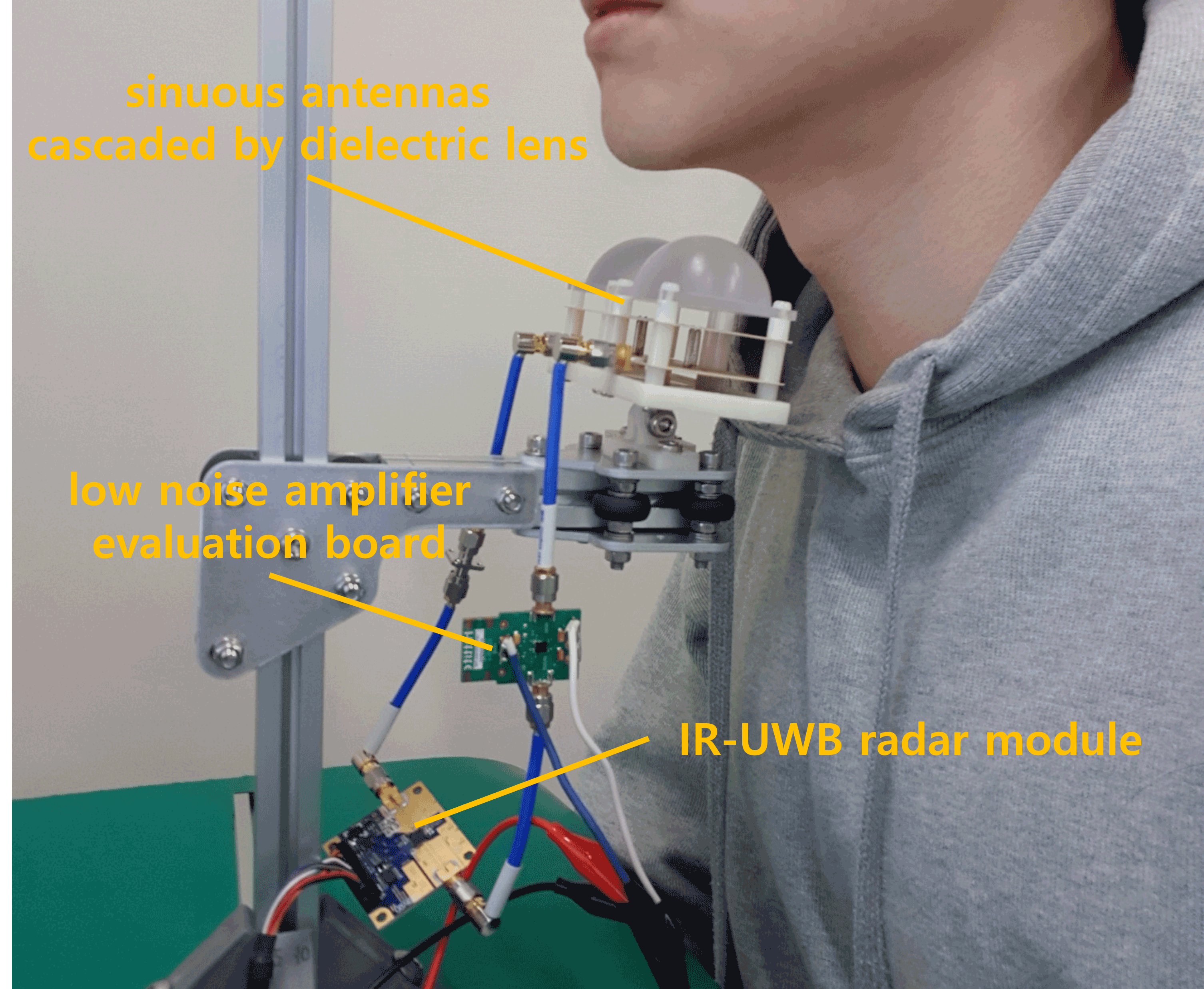}
  \caption{Experimental hardware setup for this research.}
  \label{fig:SetupPhoto}
\end{figure}

The silent speech interface (SSI), which allows speech communication without audio information, is a device that radar technology can contribute to. 
It can be an alternative HCI device for speech recognition in situations where voice-based communication is not allowed, or to people who cannot make their own voice due to disability. A myriad of sensors have been used to build SSIs \cite{Denby2010, Schultz2017}. However, many of the sensors adopted in SSI, including electromagnetic articulography \cite{Schonle1987, Kim2017}, electropalatography \cite{Woo2021}, permanent magnetic articulography \cite{Gilbert2010, Gonzalez2016}, and surface electromyography \cite{Wand2011}, operate in an invasive or contact manner, which degrades user convenience. 
However, radar sensors can operate without contact.
In addition, the impulse radio ultra-wide band (IR-UWB) radar adopted in this research has appealing properties, such as high range measurement accuracy and penetrability \cite{Khan2017}. Therefore, the IR-UWB radar has the potential to detect the movement of articulators, such as lips, jaw, and visible or invisible tongue, with high accuracy. 
In other words, IR-UWB radar has the beneficial characteristics of capturing high-quality data to classify users' various pronunciations.

In radar-based SSI, although the antennas of radar can be positioned at various places, such as in front of the mouth or face, beside the cheek, and under the chin and so forth, many of the related studies located the antennas in front of the mouth or face \cite{Eid2009, Shin2016a, Lee20191211, Wen2020, Tardif2022, Ferreira2022} to facilitate the detection of the user's lip or tongue motion. In phonetics, some pronunciations accompany fine tongue tip or tongue body motions. However, tongue movement information detected by a radar sensor whose antennas are pointed to the face or mouth can be insufficient to recognize various tongue motions in detail. Even if we use an IR-UWB radar whose transmitted signal has penetrability and antennas are pointed toward the mouth, obstacles such as skin and teeth reduce the strength of the received signal reflected from the tongue. Therefore, in addition to tongue movement information acquired from radar sensors whose antennas are pointed toward the mouth, if additional tongue movement information is collected, it will be helpful to recognize multiple pronunciations.

In this study, we tried to identify the feasibility of acquiring invisible tongue movement information by measuring the tongue and its related body parts using an IR-UWB radar, whose antennas are pointed toward the user's chin. An experiment was conducted to distinguish between the states of rest and movement of the tongue while the participants closed their mouths. We proposed a feature extraction algorithm, and used a conventional Gaussian mixture model--hidden Markov model (GMM--HMM) \cite{Gales2008} for the classification task.

\section{Methodology}

\subsection{Experimental Environment}

 In total, four participants, including two males and two females, participated in the experiment. One of the male participants was one of the authors of this study. As shown in Fig. \ref{fig:SetupPhoto}, an IR-UWB radar module with a low-noise amplifier (LNA) evaluation board and sinuous antennas cascaded with a dielectric lens was installed. The IR-UWB radar module and LNA evaluation board used in this study were NVA-R661 from Novelda and 129787-HMC902LP3E from Analog Devices, respectively. Participants were asked to sit and locate their chin approximately $5$--$10$ \SI{}{\cm} above the radar antenna.
 
 The radar sensor measured the following two states of participants: State 1 involved resting the tongue on the floor of the mouth, and state 2 involved the sequential procedure of starting from state 1, touching the palate using the tongue tip, and finishing with state 1. The participants were required to close their mouth so that the tongue was entirely invisible and not to move their body as much as possible, except for the tongue during the whole experiment. 
 The participants repeated states 1 and 2 twenty times. The starting and finishing of measuring of each state was decided by manually clicking graphic user interface button on computer screen. The data measurement period was approximately $1$--$3$ \SI{}{\second} regardless of the state.

\begin{figure} 
\centerline{\includegraphics[width=1.0\linewidth]{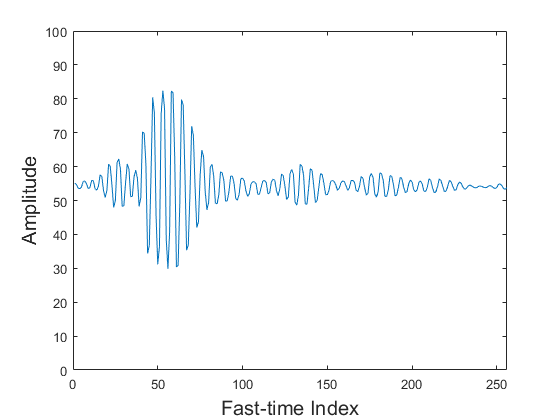}}
\caption{Example of a frame acquired from IR-UWB radar used in this research.}
\label{fig:FrameExample}
\end{figure}

 The IR-UWB radar used in this study transmits pulses to the target and receives reflected pulses from the target. 
 Then, the received pulses were merged into one frame using the manufacturer's normalization method. 
 As shown in Fig. \ref{fig:FrameExample}, one frame acquired from IR-UWB radar is sampled with 256 points called ``fast-time'' \cite{Shin2016a} index, each of which indicates distance from the radar; and each discrete index has the signal strength value which ranges from 0 to 100. 
 In the experiment, approximately 200 frames were acquired per second. 
 Every time participants repeated states 1 or 2, we collected one ``frame set'' which has the form of $T$-by-256 matrix   (i.e., one frame set consists of $T$ frames). 
 Here, $T$ is the number of acquired frames. 
 For example, if frames per second (FPS) is $200$ and the recording period of one frame set is \SI{2}{\second}, the number of acquired frames ($T$) is $400$.

\subsection{Feature Extraction and Classifier Selection}

If the IR-UWB radar had the capability to detect the motion of an invisible tongue and its related body parts, there would be a significant change in the collected radar data when the participants conducted state 2. 
However, when the participants are motionless (state 1), no significant change in the collected radar data is expected.  
Therefore, we need to devise a feature that can efficiently reflect the changes in the radar data to classify states 1 and 2.

\begin{figure}
\centerline{\includegraphics[width=0.9\linewidth]{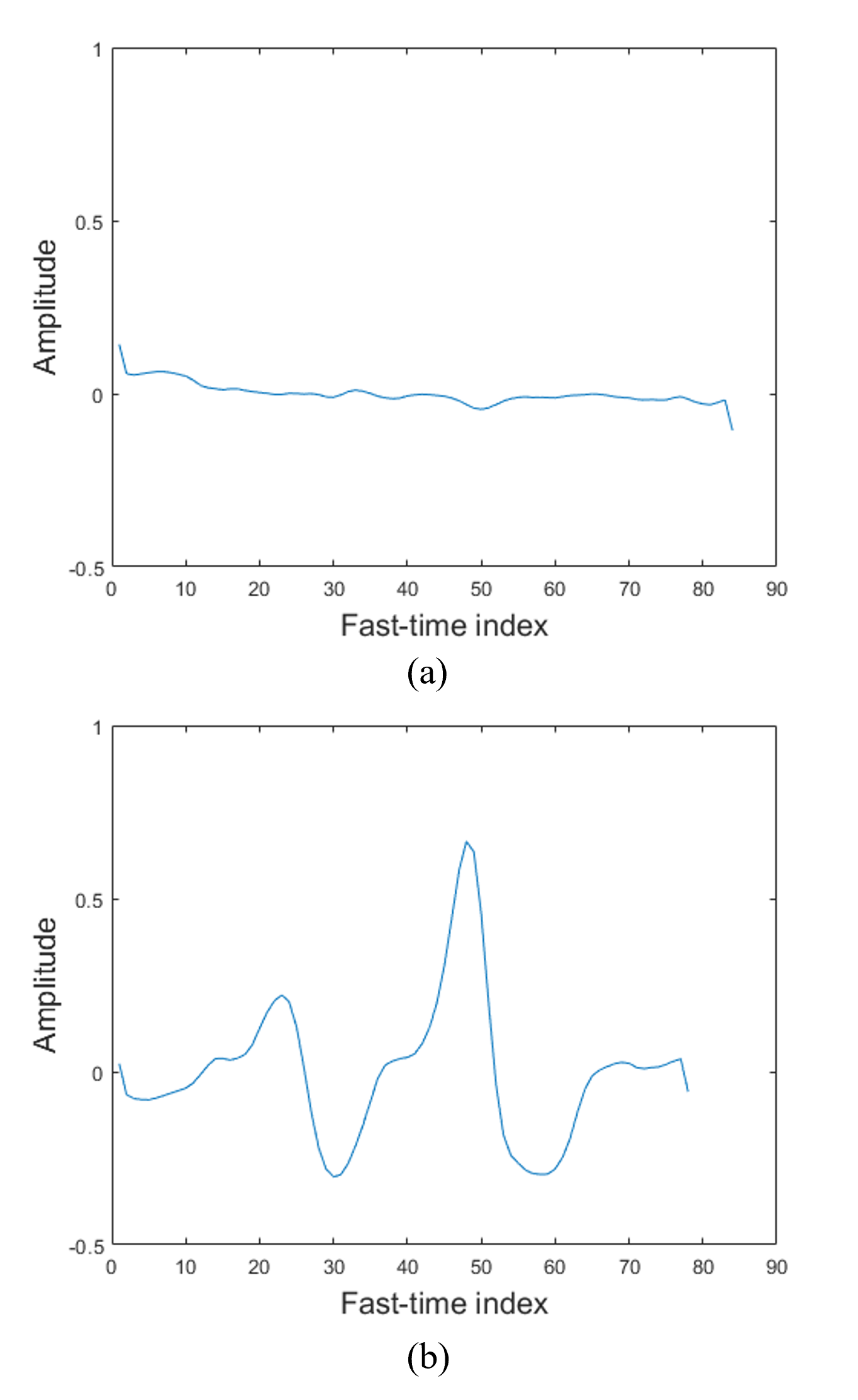}}
\caption{Acquired one-dimensional features using proposed feature extraction method from: (a) the frame set measuring a participant who did not move his tongue (state 1) and (b) the frame set measuring a participant who moved his tongue (state 2).}
\label{fig:ExtractedFeatures}
\end{figure}

We applied the following signal processing procedures to every frame set collected when the participants conducted states 1 or 2. 
First, we concatenated every frame within a frame set such that the dimension of the data became 1-by-$T \! \times \! 256$. (Remind that the dimension of a frame set is $T$-by-256.)
Then, envelope detection was performed on concatenated data.
To detect the envelope, we generated a 400-length sliding window on the concatenated data and calculated a root mean square (RMS) value of the data within each window. 
The calculated RMS values by sliding the window constitute the envelope of the concatenated data.
Finally, the detected envelope was downsampled and the mean value was subtracted such that a feature had a length of $T/4$ without a DC offset. 
As reflected in Fig. \ref{fig:ExtractedFeatures}, the feature extracted from the frame set measuring state 2 tends to show far more fluctuation than that measured in state 1.

We selected GMM--HMM \cite{Gales2008} as the classification method. It is a prevalent machine learning method for silent speech recognition task \cite{Hofe2013, Meltzner2018}.
In detail, we used a five-state left-to-right HMM and the emission probability per state was modeled with a single mixture Gaussian.
Because GMM--HMM can model sequential data with a variable length, no additional processing is required to fix the length of the feature.

\section{Evaluation Method}

To demonstrate the benefit of the proposed method, we implemented other methods and applied them to this study for comparison. 
Shin and Seo \cite{Shin2016a} proposed a short-template-based CLEAN algorithm that exhibited better performance than the conventional CLEAN algorithm, which is one of the most frequently used target detection methods in radar-based applications \cite{Kulpa2008}, in a 10-word classification task utilizing an IR-UWB radar. 
Two features, which indicate the distance of the target and strength of the radar pulse reflected from the target, were extracted at every radar frame using a conventional or short-template-based CLEAN algorithm. 
Thus, two one-dimensional variable-length features were generated for each frame set. 
The classification was performed by utilizing the multidimensional dynamic time warping (MD-DTW) algorithm, which provides the distance between two frame sets based on their multidimensional features.
Note that when implementing conventional or short-template-based CLEAN algorithm, we also applied a signal-averaging-based clutter reduction procedure, as described in \cite{Shin2016a}.

We implemented the following three types of methods and applied leave-one-out cross-validation (LOOCV) to validate the performance of each method: 
\begin{itemize}
\item Conventional CLEAN algorithm + MD-DTW 
\item Short-template-based CLEAN algorithm + MD-DTW
\item Proposed feature extraction algorithm + GMM--HMM (proposed method). 
\end{itemize}
As the participants conducted states 1 and 2 twenty times, 40 frame sets were collected per participant. 

When GMM--HMM was used for classification, two GMM--HMM models were generated: one model trained features extracted from the frame sets for state 1 and the other trained features extracted from the frame sets for state 2. 
After training, each GMM--HMM model was able to provide a probability that a given frame set would belong to the corresponding model based on the features of the given frame set. 
To apply LOOCV, each of the 40 frame sets was used for the test and the other 39 frame sets were used for training. Each test frame set was classified into the category of the GMM--HMM model that provided a larger probability between states 1 and 2.  

The MD-DTW algorithm provides the distance between two data when features from the two data are provided as input.
To classify each frame set using MD-DTW and validate the result by LOOCV, the following procedure was taken. 
We selected each of the 40 frame sets for test and obtained the distances between the test frame set and each of the other 39 frame sets using MD-DTW.
Then, the test frame set was classified into the category of the frame set that has the smallest distance to the test frame set.

\section{Results and Discussion}

We requested the participants to close their mouths so that the tongue was entirely invisible during the entire period.
When the participants moved their tongues while closing their mouths, we observed that their muscles in the throat also moved together. 
It is known that the movement of the hyoid bone or suprahyoid muscles is involved in tongue movement \cite{Hiiemae2003, Mansfield2018}. 
Thus, the movement of tongue and its related body parts is detected simultaneously by an IR-UWB radar. 

\begin{table}
\centering
\caption{Classification Accuracy (\%) for Each Individual Participants When Applying Three Types of Methods and Leave-One-Out Cross-Validation}
\label{tab:ClassificiationResult}
\begin{tabular}{|c|c|c|c|c|} 
\hline
\multirow{2}{*}{\textbf{Methods}}                                                         & \multicolumn{4}{c|}{\textbf{Participant ID}}           \\ 
\cline{2-5}
                                                                                          & \textbf{P1} & \textbf{P2} & \textbf{P3} & \textbf{P4}  \\ 
\hline
\begin{tabular}[c]{@{}c@{}}Conventional CLEAN algorithm\\+ MD-DTW\end{tabular}               & $82.5$        & $72.5$        & $82.5$        & $57.5$         \\ 
\hline
\begin{tabular}[c]{@{}c@{}}Short-template-based CLEAN algorithm\\+ MD-DTW\end{tabular}       & $85$          & $65$          & $80$          & $70$           \\ 
\hline
\begin{tabular}[c]{@{}c@{}}Proposed feature extraction algorithm \\+ GMM--HMM\end{tabular} & $100$         & $90$          & $90$          & $90$           \\
\hline
\end{tabular}
\end{table}

By utilizing three types of methods and LOOCV, we obtained the classification results for each participant in Table \ref{tab:ClassificiationResult}, where P1 (one of the authors of this paper) and P2 were male participants and P3 and P4 were female participants. Classification accuracy per participant was calculated as $(x/40)\times100$, where $x$ is the number of correctly classified frame sets. 
As summarized in Table \ref{tab:ClassificiationResult}, all the individual classification results based on the proposed method showed higher accuracy than any other method-based result.
Because the features extracted from conventional or short-template-based CLEAN algorithm only contain the information of just one arbitrary target that reflects the transmitted pulse strongest, they do not necessarily contain the movement information of the tongue and its related body parts.
Thus, conventional or short-template-based CLEAN algorithm demonstrated lower accuracy than the proposed method in the experiment. 
In this study, we identified that detecting the movement of tongue and its related body parts is feasible using an IR-UWB radar and the proposed method.

\section{Conclusion}

In this study, the classification of motionless and moving states of an invisible tongue was attempted using an IR-UWB radar whose antennas were pointed toward the chin of a participant. 
Employing the proposed method, we demonstrated a minimum classification accuracy of 90\% for four individual participants. 
Therefore, we conclude that an IR-UWB radar has the capability to detect the movement of invisible tongue and its related body parts if proper features and classification algorithms are applied. 

\section*{Acknowledgment}

This work was supported by the National Research Foundation of Korea (NRF) grant funded by the Korea government (MSIT) (NRF-2021R1F1A1062958).

\bibliographystyle{IEEEtran}
\bibliography{mybibfile, IUS_publications}

\vspace{12pt}

\end{document}